\documentclass[aps,prb,twocolumn,preprintnumbers,amsmath,amssymb,superscriptaddress]{revtex4}

\usepackage{graphicx}
\usepackage{dcolumn}
\usepackage{bm}
\usepackage{upgreek}
\begin{document}


\title{Time-Resolved Studies of a Rolled-Up Semiconductor Microtube Laser}

\author{Ch.~Strelow}
 \email{cstrelow@physnet.uni-hamburg.de}
\author{M.~Sauer}
\affiliation{%
Institut f\"{u}r Angewandte Physik und Zentrum f\"{u}r
Mikrostrukturforschung, Universit\"{a}t Hamburg, Jungiusstra{\ss}e
11, 20355 Hamburg, Germany
}%
\author{S.~Fehringer}
\author{T.~Korn}
\author{C.~Sch\"{u}ller}%
\affiliation{Institut f\"{u}r Experimentelle und Angewandte Physik,
Universit\"{a}t Regensburg, D-93040 Regensburg, Germany
}%
\author{A.~Stemmann}%
\author{Ch.~Heyn}%
\author{D.~Heitmann}%
\author{T.~Kipp\footnote{present address: Institut f\"{u}r Physikalische Chemie, Universit\"{a}t Hamburg, Grindelallee 117, 20146 Hamburg, Germany}}
\affiliation{%
Institut f\"{u}r Angewandte Physik und Zentrum f\"{u}r
Mikrostrukturforschung, Universit\"{a}t Hamburg, Jungiusstra{\ss}e
11, 20355 Hamburg, Germany
}%

\date{\today}

\begin{abstract}
We report on lasing in rolled-up microtube resonators. Time-resolved
studies on these  semiconductor lasers containing GaAs quantum wells
as optical gain material reveal particularly fast turn-on-times and
short pulse emissions above the threshold. We observe a strong
red-shift of the laser mode during the pulse emission which is
compared to the time evolution of the charge-carrier density
calculated by rate equations.

\end{abstract}

\maketitle

The most studied semiconductor microcavity lasers are micropillars
\cite{schwab06,ulrich07}, microdisks \cite{mccall92, Luo00} and
photonic crystals \cite{altug06,strauf06}. Their small size combined
with high quality factors exhibits the possibility of low threshold
lasing and single mode operation. Besides the pursuit of lowering
the threshold also fast turn-on-times are desirable
\cite{Luo00,altug06}. Presently the fastest turn-on-times were
achieved by using quantum wells (QWs) as gain material
\cite{altug06} which have the disadvantage of high non-radiative
recombination rates in the vicinity of the structured surfaces. To
overcome this problem the surface can be passivated
\cite{Mohideen94,englund07}. Quantum dots offer a much lower surface
recombination rate \cite{Luo01} but have the disadvantage of a low
carrier capture rate \cite{ellis07} which lengthens the
turn-on-time. In this work we present lasing in microtubes. These
structures have the intrinsical advantage of nearly no unpassivated
surfaces. Exploiting the self-rolling mechanism of thin strained
semiconductor bilayers \cite{Prinz00} we fabricated multi-walled
microtubes that act as optical ring resonators \cite{Kipp06}. As
optical gain medium we chose a GaAs QW. Applying sub-picosecond
optical pumping we measured the time evolution of the
photoluminescence (PL) emission at low temperature.

\begin{figure}[b]
\includegraphics[width=7.6cm]{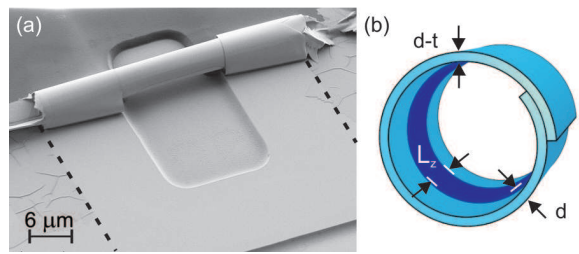}
\caption{\label{fig1} (a) Scanning electron micrograph a microtube and its U-shaped mesa (dashed lines).
(b) Unscaled sketch of the free-standing part of the tube.}
\end{figure}

Our samples were grown by molecular beam epitaxy on
$\langle$001$\rangle$ GaAs substrates. On top of a 40 nm AlAs
sacrificial layer a 42 nm thick layer system is grown consisting of
a 16 nm thick pseudomorphically strained barrier layer of
In$_{0.16}$Al$_{0.30}$Ga$_{0.54}$As, a 4 nm GaAs QW, and a 22 nm
Al$_{0.20}$Ga$_{0.80}$As barrier. By selectively etching the
sacrificial layer a predefined U-shaped mesa rolls up and forms a
free-standing microtube bridge. Details of the fabrication are
described in Refs. \onlinecite{Kipp06, Strelow07}. The tube
investigated in this work is shown in Fig.\,\ref{fig1}(a). It has a
diameter of 5.74 $\upmu$m and 1.1 revolutions. In the free-standing
part total internal reflection leads to waveguiding in the tube wall
and to the formation of ring modes by constructive interference
after a roundtrip \cite{Kipp06}. Very recently we showed that a
fully three-dimensional control of the eigenmodes can be achieved by
the definition of lobes in the rolling edge \cite{Strelow08}. In
this work we used a different axial confinement mechanism. By
etching two $t$=4 nm deep stripes of a distance $L_z$=1.43 $\upmu$m
before rolling-up we fabricated a rolled-up ridge waveguide, as
shown schematically in Fig.\,\ref{fig1}(b). Here, it also becomes
obvious that microtubes should have low surface recombination rates
due to the nearly complete absence of gain material at the surface.

Our experiments were performed with a tunable Ti:Sa laser with 600
fs pulse length and 81 MHz repetition rate. The laser was focussed
by a 40$\times$ microscope objective (NA=0.4) onto the sample
mounted in a continuous flow He cryostat at 4 K.  The PL light is
collected by the same objective, dispersed by a 25 cm spectrometer
with 600 lines/mm grating  and detected either by a streak camera
(Hamamatsu Synchroscan) or a Si charge coupled device camera. The
time resolution of the setup is 9 ps, measured from the FWHM of the
pump laser. The excitation wavelength was set to 750 nm which
excites only the QW and not the barriers, since the band gap of the
InAlGaAs layer is increased by compressive strain inside the
microtube structure.
\begin{figure}[t]
\includegraphics[width=7.4cm]{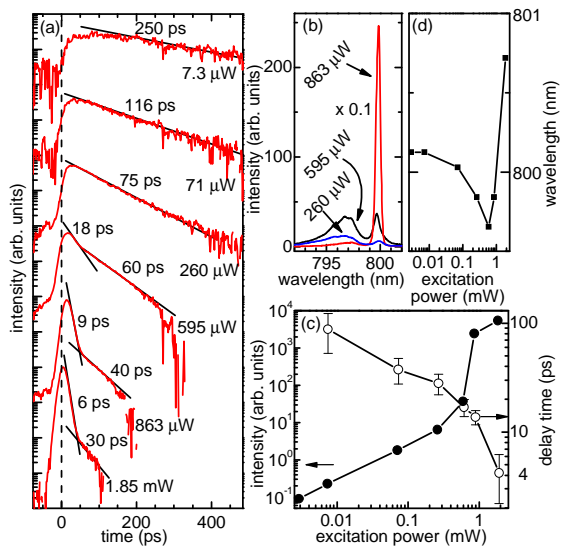}
\caption{\label{fig2} Excitation power dependent measurements of the
microtube in Fig.\,\ref{fig1}. (a) Time evolution of the emitted PL light.
(b) PL spectra. (c) Time-integrated PL intensity and delay time between pump
 pulse and PL emission. (d) Wavelength of the lasing mode in the
 time-integrated spectra.}
\end{figure}
Figure \ref{fig2} shows excitation power dependent measurements on
our microtube. Emission spectra of the microtube for three different
excitation powers (measured as time-averaged power of the pulsed
laser) are shown in Fig.\,\ref{fig2}(b). The QW emits around 797 nm
with a FWHM of about 4.5 nm. Its emission is modified by optical
modes. We rule out strong-coupling effects since all estimated
charge carrier densities exceed the saturation density for strong
coupling\cite{Houdre95}. The strong mode at 800 nm is an axial
fundamental mode \cite{Strelow08} with one antinode in axial
direction, determined from spatially resolved measurements (not
shown). Figure \ref{fig2}(a) depicts the time evolution of the PL
emission of the mode for different excitation powers. With
increasing excitation power one clearly observes a shortening of the
lifetime from 250 ps down to about 6 ps, determined by fitting the
decay in a period longer than the time resolution (at least 20 ps)
and calculating the 1/e lifetime. In addition we find a threshold
behavior in the light-input/light-ouput (L-L) curve depicted in
Fig.\,\ref{fig2}(c). Above a threshold between 260 $\upmu$W and 59
$\upmu$W the system starts lasing. With increasing excitation power
more and more stimulated emission shortens the decay times and a
short pulse appears at the onset of the PL emission. For 863
$\upmu$W a single laser mode dominates the spectrum, as shown in
Fig.\,\ref{fig2}(b). Interestingly, the decay times already shorten
well below the threshold. We believe that between 7.3 $\upmu$W and
71 $\upmu$W population inversion is reached and amplified
spontaneous emission shortens the decay time. The threshold power is
quite high compared to values for microdisk lasers \cite{Luo01,
zwiller03} and photonic crystal laser \cite{altug06, ellis07}. This
is caused by the large surface to volume ratio of microtubes. In
finite-difference time-domain simulations, assuming that only the
quantum well absorbs power, we calculated that less than 2 \% of the
laser power are absorbed by the microtube.  The threshold power
might be much lower if one overcomes the difficulties of optical
pumping, e.g., by electrical pumping.

Figure \ref{fig2}(c) depicts the delay between the arrival of the
pump pulse, determined by the stray light of the excitation laser,
and the maximum of the PL emission. A clear drop of the
turn-on-times with increasing excitation power is observed. For 863
$\upmu$W we measured about 14 ps. By exciting with light linearly
polarized perpendicular to the mode
polarization\cite{Kipp06,Strelow07} and measuring in corresponding
polarizations, turn-on-times even below the conventional time
resolution can be determined. For 1.85 mW we find $4 \pm 1.5$ ps.
Our turn-on-times are quite fast in comparison to values in Refs.
\onlinecite{altug06,Luo01}, despite the absence of a fastened
spontaneous decay rate by the Purcell effect. We attribute this to
the nearly-resonant carrier excitation directly in the QW and the
fast carrier relaxation in the QW. In addition we observed no
relaxation oscillations in our experiments as often observed in
microdisks \cite{Luo00,Luo01}. This is because microtubes have no
charge-carrier reservoir like microdisks. In QW microdisk lasers
diffusion from the middle part to the active region is a
prerequisite in order to balance gain losses due to the very strong
non-radiative surface state recombination.

\begin{figure}[t]
\includegraphics[width=7.50cm]{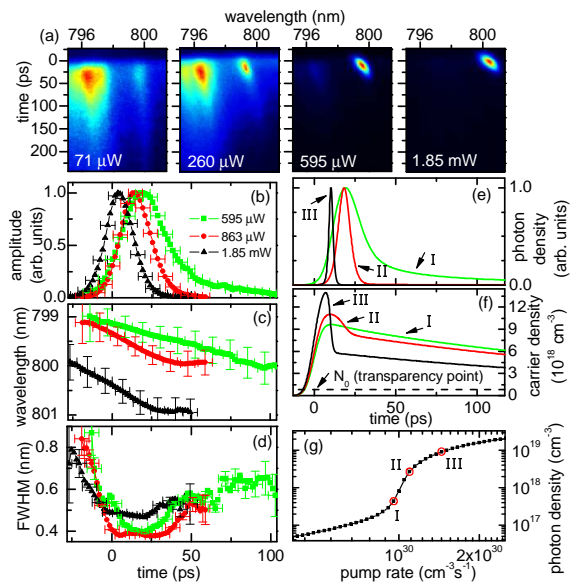}
\caption{\label{fig3} (a) False-color images by the streak camera for different excitation powers.
(b)-(d) Time evolution of the amplitude, wavelength and width of the lasing mode obtained by
Gaussian fits at all times in (a). (e)-(g) Results for the time evolution of the photon density (e) and
the charge carrier density (f) for three pump rates marked in the pump rate dependent photon density in (g) calculated by
rate equations.}
\end{figure}

Figure \ref{fig2}(d) shows the center wavelength of the time
integrated lasing mode in dependence on the excitation power. Up to
595 $\upmu$W we observe a strong blue shift followed by a strong red
shift for higher excitation powers. The blue shift is caused by a
change of the refractive index with an increasing charge carrier
density for increasing excitation powers. Although untypical for
pulsed measurements we attribute the red shift to heating which even
destroys microtubes when excited with powers of typically 4 mW.
Probably the thin microtube walls offer only a slow heat transfer to
the substrate. A wavelength shift is also observed during the short
PL emission and gives insight into the charge carrier dynamics on
short time scales: Figure \ref{fig3} (a) depicts false-color images
taken by the streak camera at different excitation powers.
Obviously, the lasing mode around 800 nm shifts to red during the
emitted light pulse and the shift becomes larger for higher
excitation powers. We fitted the spectral profiles at different
times with Gaussians for the highest three excitation powers in
Fig.\,\ref{fig3}(a). The results in form of amplitude, wavelength
and width are depicted in Fig.\,\ref{fig3}(b)-(d). Interestingly, we
also observe a dynamic change of the spectral width besides the red
shift. The first photons emitted after excitation are emitted
spontaneously and the line width reflects the cavity life time.
Indeed, the width of 0.8 nm at the onset of PL emission is the same
as measured well below the threshold in the time-integrated spectra.
Then, the width decreases by 50\% within about 20 ps, stays constant
for 30 ps (when the maximum of the PL emission is reached) and again
increases at the end of the PL pulse. We attribute this behavior to
the dynamic change into the coherent light state, when the system
lases, and back into the spontaneous emission regime when the lasing
turns off. Line-width narrowing with increasing pump power is
reported for other microlasers \cite{strauf06}, but here we observe
it on the time scale.

To explain the time evolution of the emitted PL intensity and the
wavelength of the lasing mode we numerically solved rate equations
for the charge-carrier density $N$ and the photon density $S$:
\begin{eqnarray}
\dot{N}&=&-\Gamma S G_0 (N-N_0)-\frac{N}{\tau_{sp}}+P(t),\label{1}\\
\dot{S}&=&\Gamma S G_0 (N-N_0)+\beta \frac{N}{\tau_{sp}}-\frac{S}{ \tau_p}.\label{2}
\end{eqnarray}
Here, we assumed a constant differential gain of $G_0=5\times
10^{-6}$ cm$^3/$s and a value of $N_0=1.16\times 10^{18}$ cm$^{-3}$
for $N$ at the transparency point.\cite{Luo00} For the optical
confinement factor $\Gamma$ we calculated $0.0563$ and for the
spontaneous emission factor $\beta$ we assumed 0.01. According to
the experiment we chose the spontaneous life time to $\tau_{sp}=250$
ps [see Fig.\,2(a) for low excitation power]. We neglected
non-radiative decays since for our comparative studies significant
non-radiative decay rates do not change the results qualitatively.
From the experimentally obtained cavity quality factor $Q=1000$ we
determined a photon lifetime $\tau_p$ in the cavity of $0.424$ ps.
The pump pulse we assumed to be Gaussian, i.e., $P(t)=P_0
e^{-(2t/w\sqrt{ln\,2})^2}$ with a FWHM of $w=$10 ps. The
time-integrated photon density versus the pump rate is shown in
Fig.\,\ref{fig3} (g). A qualitative agreement to the experimental
data in Fig.\,\ref{fig2} (c) is observed. Figures\,\ref{fig3} (e)
and (f) depict the time evolution of $S$ (normalized) and $N$ for
pump rates roughly corresponding to the excitation powers in the
Fig.\,\ref{fig3} (b)-(d). The calculated curve of $S$ qualitatively
reproduces the experimental data in Fig.\,\ref{fig3}(b). The time
evolution of $N$ can be compared to the time evolution of the
wavelength in Fig.\,\ref{fig3}(c). $N$ changes the refractive index
linearly \cite{Mendoza80} which on the other hand causes a linear
shift of the mode wavelength. Consequently, the wavelength shift is
a direct measure for $N$. Slightly above the threshold (pump rate I)
$N$ decays mainly spontaneously and thus exponentially. Well above
the threshold (pump rates II, III) dominating stimulated emission
leads to the observed fast drop of $N$. This behavior can directly
be found in the measurements: At 595 $\upmu$W (slightly above the
threshold)the wavelength shift is slow and at 863 $\upmu$W (well
above the threshold) it changes from fast during the pulse emission
to slow at its end. For 1.85 mW the shift does not become faster in
contrast to the calculations. We believe that here the assumption of
a constant gain breaks down. The mode might have shifted to
different gain or the gain itself has changed e.g. due to saturation
or renormalization effects or heating. The absolute values of the
wavelength are generally not reproduced by the calculations. For
high excitation powers heating leads to an additional red-shift of
the laser mode.

In summary, we report on the demonstration of lasing in active
microtube resonators. We observed fast turn-on times as compared to
other micro-lasers and a characteristic red shift of the laser mode
during pulsed emission. This behavior is reproduced by a rate
equation model.

We acknowledge financial support by the Deutsche
Forschungsgemeinschaft via SFB 508 and Graduiertenkolleg 1286 and
638.

%
%

\end{document}